\documentclass[preprint,eqsecnum,prd,nofootinbib]{revtex4}   
\usepackage{epsf}
\usepackage{graphicx}

\def\beq{\begin{equation}}
\def\eeq{\end{equation}}
\def\beqa{\begin{eqnarray}}
\def\eeqa{\end{eqnarray}}

\newlength{\dinwidth} \newlength{\dinmargin}
\begin{document}
\preprint{Cavendish-HEP-03/15}

\preprint{FSU-HEP-030725}

\title{Next-to-next-to-leading-order soft-gluon corrections \\
in direct photon production}
\author{Nikolaos Kidonakis}
\affiliation{ Cavendish Laboratory, University of Cambridge,\\
Madingley Road, Cambridge CB3 0HE, UK}
\author{J.F. Owens}
\affiliation{ Physics Department, Florida State University,\\ 
Tallahassee, FL 32306-4350, USA}

\begin{abstract}
Previous work on soft-gluon resummation for direct photon production 
is extended to include additional subleading logarithmic terms through 
${\cal O}(\alpha \alpha_s^3)$ and some representative comparisons are made to 
experimental results from the E-706 and UA-6 Collaborations. The additional 
terms are small in magnitude, indicating good convergence properties to the 
level of accuracy calculated. The scale dependence remains much smaller than 
that of the next-to-leading-order calculation.
\end{abstract}
\maketitle

\section{Introduction}
\label{sec:intro}

Direct photon production is widely recognized as a process that is potentially 
important in determinations of the gluon distribution function.
The next-to-leading-order (NLO) cross section for direct photon production 
has been given in Refs. \cite{ADBFS,GV}. The role of higher-order soft-gluon
corrections has also been addressed more recently.
Threshold resummation studies for direct photon production
have appeared in Refs. \cite{LOS,CMNOV,KO1,StVog} while a joint 
threshold and transverse momentum resummation formalism 
has been given in Ref. \cite{LSV}. 

In a previous paper \cite{KO1} we presented analytical and numerical results
for the next-to-next-to-leading-order 
(NNLO) next-to-next-to-leading-logarithm (NNLL) soft-gluon corrections
for direct photon production.
Here we increase the accuracy of our previous calculation 
by including additional subleading soft corrections. Our approach follows 
Ref. \cite{NKNNLO} which in turn is based on and extends
previous work on threshold resummation \cite{LOS,KS,KOS,KV,NK,KO2}.

\section{NNLO soft corrections}
\label{sec:calculation}
At lowest order, the parton-parton scattering  subprocesses are 
$ q(p_a)+g(p_b) \rightarrow  \gamma(p_{\gamma}) + q(p_J)$
and $q(p_a)+{\bar q}(p_b) \rightarrow \gamma(p_{\gamma}) + g(p_J)$. 
We define the Mandelstam invariants 
$s=(p_a+p_b)^2$, $t=(p_a-p_{\gamma})^2$, and 
$u=(p_b-p_{\gamma})^2$,
which satisfy $s_4 \equiv s+t+u=0$ at threshold.
Note that the photon transverse momentum is
$p_T=(tu/s)^{1/2}$. Here we calculate the 
cross section $E_{\gamma} \; d^3\sigma/d^3 p_{\gamma}$
in single-particle-inclusive
kinematics in the $\overline {\rm MS}$ scheme.
The soft corrections to the cross section appear in the form of 
plus distributions
\beq
{\cal D}_l(s_4)\equiv\left[\frac{\ln^l(s_4/p_T^2)}{s_4}\right]_+
\eeq
with $l \le 2n-1$ at $n$th order in $\alpha_s$ beyond the leading order,
while the virtual corrections appear in $\delta(s_4)$ terms.

We begin with the NLO soft and virtual corrections in the 
$\overline{\rm MS}$ scheme. A somewhat different notation from that 
used in Ref. \cite{KO1} has been 
adopted here, so the previously calculated terms are repeated here, as well. 
The corrections to the parton-level cross section, ${\hat \sigma}$, 
can be written for either subprocess as
\beq
E_{\gamma} \frac{d^3{\hat\sigma}^{(1)}_{f_i f_j}}
{d^3 p_{\gamma}}=\sigma^B_{f_i f_j}
\frac{\alpha_s(\mu_R^2)}{\pi} \left\{c_3^{f_i f_j} \, {\cal D}_1(s_4)
+c_2^{f_i f_j} \,  {\cal D}_0(s_4) + c_1^{f_i f_j} \,  \delta(s_4) 
\right\}\, ,
\eeq
where $\mu_R$ is the renormalization scale, and the Born terms are given by
\beq
\sigma^B_{q{\bar q}}
=\frac{2 C_F}{N_c}\frac{\alpha \alpha_s}{s} e_q^2
\left(\frac{t}{u}+\frac{u}{t}\right)\, , \quad
\sigma^B_{qg}= 
-\frac{1}{N_c}\frac{\alpha \alpha_s}{s} e_q^2 
\left(\frac{s}{t}+\frac{t}{s}\right) \, ,
\eeq
where $e_q$ is the charge of a quark of type $q$, and
$C_F=(N_c^2-1)/(2N_c)$ with $N_c=3$ the number of colors.
Also 
$c_3^{q{\bar q}}=4 C_F-C_A$, $c_3^{qg}=C_F+2C_A$,
\beq
c_2^{q{\bar q}}=-\frac{\beta_0}{4}
-2C_F \ln\left(\frac{\mu_F^2}{p_T^2}\right)\, , \quad
c_2^{qg}=-\frac{3}{4}C_F-(C_F+C_A)\ln\left(\frac{\mu_F^2}{p_T^2}\right)\, ,
\eeq
where $\mu_F$ is the factorization scale, $C_A=N_c$, and
$\beta_0=(11C_A-2n_f)/3$, with $n_f$ the number of quark flavors.
We also define for use below 
$T_2^{q{\bar q}}=-(\beta_0/4)
+2C_F \ln(p_T^2/s)$ and 
$T_2^{qg}=-(3/4) C_F
+(C_F+C_A) \ln(p_T^2/s)$.
Finally we write
$c_1^{f_i f_j}=c_1^{\mu \;{f_i f_j}}+T_1^{f_i f_j}$.
For $q{\bar q} \rightarrow \gamma g$ we have
\beq
c_1^{\mu \;{q{\bar q}}}=C_F\left[-\frac{3}{2}
-\ln\left(\frac{p_T^2}{s}\right)\right]
\ln\left(\frac{\mu_F^2}{s}\right) 
+\frac{\beta_0}{4} \ln\left(\frac{\mu_R^2}{s}\right),
\eeq
and $T_1^{q{\bar q}}=C_F[3/2
+\ln(p_T^2/s)]\ln(p_T^2/s) -(\beta_0/4)\ln(\mu_R^2/s)
+{c'}_1^{q \bar q}$ where ${c'}_1^{q \bar q}$ is defined in 
Eq. (3.11) of Ref. \cite{KO1}.
For $qg \rightarrow \gamma q$ we have
\beq
c_1^{\mu \; {qg}}=\left[-\frac{\beta_0}{4}-\frac{3}{4}C_F
-C_F\ln\left(\frac{-t}{s}\right)-C_A\ln\left(\frac{-u}{s}\right)\right]
\ln\left(\frac{\mu_F^2}{s}\right) 
+\frac{\beta_0}{4} \ln\left(\frac{\mu_R^2}{s}\right)
\eeq
and $T_1^{qg}=[3C_F/4+C_F\ln(-t/s)+C_A\ln(-u/s)]
\ln(p_T^2/s)-(\beta_0/4)\ln(\mu_R^2/p_T^2)
+{c'}_1^{qg}$ where ${c'}_1^{qg}$ is defined in 
Eq. (3.8) of Ref. \cite{KO1}.

Note that the NLO $c_i$ coefficients have also been presented in 
Ref. \cite{KO1}.
The notation for  $c_3$ and $c_2$  is the same as before,
while the notation for splitting $c_1$ into $c_1^{\mu}$ and $T_1$ terms 
for each subprocess is new and useful in presenting the NNLO expressions 
below.

Next, we turn to the NNLO soft and virtual corrections 
in the $\overline{\rm MS}$ scheme.
These corrections can be written for either channel as 
\beq
E_{\gamma} \frac{d^3{\hat\sigma}^{(2)}_{f_i f_j}}
{d^3 p_{\gamma}}=\sigma^B_{f_i f_j}
 \frac{\alpha_s^2(\mu_R^2)}{\pi^2} \, 
{\hat{\sigma'}}^{(2)}_{f_i f_j} \, .
\label{NNLOm}
\eeq
For the $q{\bar q}\rightarrow \gamma g$ process we have
\beqa
{\hat{\sigma'}}^{(2)}_{q{\bar q}}&=& 
\frac{1}{2} (c_3^{q{\bar q}})^2 \, \,  {\cal D}_3(s_4)
+\left[\frac{3}{2} c_3^{q{\bar q}} \, c_2^{q{\bar q}} 
- \frac{\beta_0}{4} c_3^{q{\bar q}}
+C_A \frac{\beta_0}{8}\right] \, {\cal D}_2(s_4)
\nonumber \\ && \hspace{-10mm}
{}+\left\{c_3^{q{\bar q}} \, c_1^{q{\bar q}} +(c_2^{q{\bar q}})^2
-\zeta_2 \, (c_3^{q{\bar q}})^2 -\frac{\beta_0}{2} \, T_2 ^{q{\bar q}}
+\frac{\beta_0}{4} c_3^{q{\bar q}}  \ln\left(\frac{\mu_R^2}{s}\right)
+2 \, C_F \, K \right.
\nonumber \\ && \hspace{-10mm} \quad \quad \left.
{}+C_A \left[-\frac{K}{2} 
+\frac{\beta_0}{4} \, \ln\left(\frac{p_T^2}{s}\right)\right]
-\frac{\beta_0^2}{16} \right\} \, {\cal D}_1(s_4)
\nonumber \\ && \hspace{-10mm} 
{}+\left\{c_2^{q{\bar q}} \, c_1^{q{\bar q}} 
-\zeta_2 \, c_2^{q{\bar q}} \, c_3^{q{\bar q}}
+\zeta_3 \, (c_3^{q{\bar q}})^2 
-\frac{\beta_0}{2} \, T_1^{q{\bar q}}
+\frac{\beta_0}{4}\, c_2^{q{\bar q}} \ln\left(\frac{\mu_R^2}{s}\right) 
+G^{(2)}_{q {\bar q}}
\right. 
\nonumber \\ && \hspace{-10mm} \quad \quad
{}+C_F \left[\frac{\beta_0}{4} 
\ln^2\left(\frac{\mu_F^2}{s}\right)
- K \, \ln\left(\frac{\mu_F^2}{s}\right)
+K \,  \ln\left(\frac{p_T^2}{s}\right)\right]
\nonumber \\ && \hspace{-10mm} \quad \quad \left.
{}+C_A \, \left[\frac{\beta_0}{8}
\ln^2\left(\frac{p_T^2}{s}\right)
-\frac{K}{2}\ln\left(\frac{p_T^2}{s}\right)\right]
-\frac{\beta_0^2}{16} \ln\left(\frac{p_T^2}{s}\right) \right\} \, 
{\cal D}_0(s_4)  
\nonumber \\ &&  \hspace{-10mm}
{}+ R^{q{\bar q}\rightarrow \gamma g} \, \, \delta(s_4) \, .
\label{NNLOqqbar}
\eeqa
Here $K= C_A\; (67/18-\pi^2/6) - 5n_f/9$, 
$\zeta_2=\pi^2/6$, and $\zeta_3=1.2020569\cdots$.
The function $G^{(2)}_{q {\bar q}}$ denotes 
a set of two-loop contributions \cite{NKNNLO} 
and is given by
\beq
G^{(2)}_{q {\bar q}}
=C_F C_A \left(\frac{7}{2} \zeta_3
+\frac{22}{3}\zeta_2-\frac{299}{27}\right)
+n_f C_F \left(-\frac{4}{3}\zeta_2+\frac{50}{27}\right) \, . 
\eeq
We determine in the virtual corrections $R^{q{\bar q}\rightarrow \gamma g}$ 
only the terms that involve the renormalization and factorization scales,
denoted as $R^{\mu \, q{\bar q}\rightarrow \gamma g}$ and given explicitly by
\beqa
R^{\mu \, q{\bar q}\rightarrow \gamma g}&=& 
\ln^2\left(\frac{\mu_F^2}{p_T^2}\right)
\left\{\frac{C_F^2}{2}\left[\frac{3}{2}
+\ln\left(\frac{p_T^2}{s}\right)\right]^2-2 \zeta_2 C_F^2
+\frac{3}{16} \beta_0 C_F+ \frac{\beta_0}{8} C_F 
\ln\left(\frac{p_T^2}{s}\right)\right\}
\nonumber \\ && \hspace{-5mm}
{}+\ln\left(\frac{\mu_F^2}{p_T^2}\right)
\ln\left(\frac{\mu_R^2}{p_T^2}\right)
\frac{(-\beta_0)}{2}C_F\left[\frac{3}{2}
+\ln\left(\frac{p_T^2}{s}\right)\right]
+\ln^2 \left(\frac{\mu_R^2}{p_T^2}\right) \, \frac{\beta_0^2}{16}
\nonumber \\ && \hspace{-5mm}
{}+\ln\left(\frac{\mu_F^2}{p_T^2}\right)
\left\{C_F^2 \left[\frac{3}{2}
+\ln\left(\frac{p_T^2}{s}\right)\right]^2 \ln\left(\frac{p_T^2}{s}\right)
-C_F \left[\frac{3}{2}+\ln\left(\frac{p_T^2}{s}\right)\right] 
\left[\frac{\beta_0}{4} \ln\left(\frac{p_T^2}{s}\right)
+T_1^{q{\bar q}} \right] \right.
\nonumber \\ && \hspace{20mm}
{}-\frac{\zeta_2}{2} \beta_0 C_F -C_F \frac{K}{2} 
\ln\left(\frac{p_T^2}{s}\right)
+C_F^2\left(-11 \zeta_3 + \frac{3}{2} \zeta_2 -\frac{3}{16}\right)
\nonumber \\ && \hspace{20mm} \left.
{}+C_F C_A \left(\frac{7}{2} \zeta_3-\frac{11}{6}\zeta_2
-\frac{17}{48}\right) + n_f C_F \left(\frac{\zeta_2}{3}
+\frac{1}{24}\right)\right\}
\nonumber \\ && \hspace{-5mm}
{}+\ln\left(\frac{\mu_R^2}{p_T^2}\right) \left\{-\frac{\beta_0}{2}C_F
\left[\frac{3}{2}+\ln\left(\frac{p_T^2}{s}\right)\right]  
\ln\left(\frac{p_T^2}{s}\right)+\frac{\beta_0}{2} T_1^{q{\bar q}}
+\frac{\beta_0^2}{8} \ln\left(\frac{p_T^2}{s}\right) 
+\frac{\beta_1}{16}\right\} \, ,
\eeqa
where $\beta_1=34 C_A^2/3 \, - \, 2n_f(C_F+5C_A/3)$ and
\beq
{\gamma'}_{q/q}^{(2)}=C_F^2\left(\frac{3}{32}-\frac{3}{4}\zeta_2
+\frac{3}{2}\zeta_3\right)
+C_F C_A\left(-\frac{3}{4}\zeta_3+\frac{11}{12}\zeta_2+\frac{17}{96}\right)
+n_f C_F \left(-\frac{\zeta_2}{6}-\frac{1}{48}\right)\, .
\eeq

For the $q g \rightarrow \gamma q$ process we have
\beqa
{\hat{\sigma'}}^{(2)}_{q g}&=& 
\frac{1}{2} (c_3^{qg})^2 \, \, {\cal D}_3(s_4)
+\left[\frac{3}{2} c_3^{qg} \, c_2^{qg} 
- \frac{\beta_0}{4} c_3^{qg}
+C_F \frac{\beta_0}{8}\right] \, {\cal D}_2(s_4)
\nonumber \\ && \hspace{-10mm}
{}+\left\{c_3^{qg} \, c_1^{qg} +(c_2^{qg})^2
-\zeta_2 \, (c_3^{qg})^2 -\frac{\beta_0}{2} \, T_2 ^{qg}
+\frac{\beta_0}{4} c_3^{qg}  \ln\left(\frac{\mu_R^2}{s}\right)
+(C_F+C_A) \, K \right.
\nonumber \\ && \hspace{-10mm} \quad \quad \left.
{}+C_F \left[-\frac{K}{2} 
+\frac{\beta_0}{4} \, \ln\left(\frac{p_T^2}{s}\right)\right]
-\frac{3}{16} \beta_0 C_F \right\} \,
{\cal D}_1(s_4)
\nonumber \\ && \hspace{-10mm} 
{}+\left\{c_2^{qg} \, c_1^{qg} 
-\zeta_2 \, c_2^{qg} \, c_3^{qg}
+\zeta_3 \, (c_3^{qg})^2 
-\frac{\beta_0}{2} T_1^{qg}
+\frac{\beta_0}{4}\, c_2^{qg} \ln\left(\frac{\mu_R^2}{s}\right) 
+G^{(2)}_{qg}
\right. 
\nonumber \\ && \hspace{-10mm} \quad \quad
{}+(C_F+C_A) \left[\frac{\beta_0}{8} 
\ln^2\left(\frac{\mu_F^2}{s}\right)
- \frac{K}{2} \, \ln\left(\frac{\mu_F^2}{s}\right)\right]
+C_F\,  K \,  \ln\left(\frac{-t}{s}\right)
+C_A\,  K \,  \ln\left(\frac{-u}{s}\right)
\nonumber \\ && \hspace{-10mm} \quad \quad \left.
{}+C_F \, \left[\frac{\beta_0}{8}
\ln^2\left(\frac{p_T^2}{s}\right)
-\frac{K}{2}\ln\left(\frac{p_T^2}{s}\right)\right]
-\frac{3}{16} C_F \beta_0 \ln\left(\frac{p_T^2}{s}\right) \right\} \,
{\cal D}_0(s_4)  
\nonumber \\ &&  \hspace{-10mm}
{}+ R^{qg\rightarrow \gamma q} \, \, \delta(s_4) \, .
\label{NNLOqg}
\eeqa
The function $G^{(2)}_{q g}$ denotes a set of two-loop contributions 
\cite{NKNNLO} and is given by
\beqa
G^{(2)}_{q g}&=&C_F^2\left(-\frac{3}{32}+\frac{3}{4}\zeta_2
-\frac{3}{2}\zeta_3\right)+ C_F C_A \left(\frac{3}{4} \zeta_3
-\frac{11}{12}\zeta_2-\frac{189}{32}\right)
\nonumber \\ && \hspace{-5mm}
{}+C_A^2 \left(\frac{7}{4} \zeta_3
+\frac{11}{3}\zeta_2-\frac{41}{216}\right)
+ n_f C_F \left(\frac{1}{6}\zeta_2+\frac{17}{16}\right) 
+n_f C_A \left(-\frac{2}{3}\zeta_2
-\frac{5}{108}\right) \, .
\eeqa
Finally, the terms in $R^{qg\rightarrow \gamma q}$
that involve the renormalization and factorization scales,
denoted as $R^{\mu \, qg\rightarrow \gamma q}$, are given explicitly by
\beqa
R^{\mu \, qg\rightarrow \gamma q}&=&\ln^2\left(\frac{\mu_F^2}{p_T^2}\right)
\left\{\frac{1}{2}\left[\frac{\beta_0}{4}+\frac{3}{4} C_F
+C_F\ln\left(\frac{-t}{s}\right)+C_A\ln\left(\frac{-u}{s}\right)
\right]^2- \frac{\zeta_2}{2} (C_F+C_A)^2 \right.
\nonumber \\ && \hspace{20mm} \left.
{}+\frac{\beta_0}{8} \left[\frac{\beta_0}{4}+\frac{3}{4}C_F
+C_F\ln\left(\frac{-t}{s}\right)
+C_A\ln\left(\frac{-u}{s}\right)\right]\right\}
\nonumber \\ && \hspace{-10mm}
{}+\ln\left(\frac{\mu_F^2}{p_T^2}\right)
\ln\left(\frac{\mu_R^2}{p_T^2}\right)
\frac{(-\beta_0)}{2}\left[\frac{\beta_0}{4}+\frac{3}{4}C_F
+C_F\ln\left(\frac{-t}{s}\right)
+C_A\ln\left(\frac{-u}{s}\right)\right]
+\ln^2 \left(\frac{\mu_R^2}{p_T^2}\right) \, \frac{\beta_0^2}{16}
\nonumber \\ && \hspace{-10mm}
{}+\ln\left(\frac{\mu_F^2}{p_T^2}\right)
\left\{\left[\frac{\beta_0}{4}+\frac{3}{4}C_F
+C_F\ln\left(\frac{-t}{s}\right)
+C_A\ln\left(\frac{-u}{s}\right)\right]^2 \ln\left(\frac{p_T^2}{s}\right)
\right.
\nonumber \\ && \hspace{15mm}
{}-\left[\frac{\beta_0}{4}+\frac{3}{4}C_F
+C_F\ln\left(\frac{-t}{s}\right)
+C_A\ln\left(\frac{-u}{s}\right)\right]
\left[\frac{\beta_0}{4} \ln\left(\frac{p_T^2}{s}\right)
+T_1^{qg} \right]
\nonumber \\ && \hspace{15mm}
{}-\frac{K}{2}\left[C_F \ln\left(\frac{-t}{s}\right)
+C_A \ln\left(\frac{-u}{s}\right)\right]
-C_F^2\left(\frac{5}{2} \zeta_3 + \frac{3}{32}\right)
-C_A^2\left(\frac{11}{4}\zeta_3+\frac{2}{3}\right)
\nonumber \\ && \hspace{15mm} \left.
{}-C_F C_A \left(\frac{9}{4} \zeta_3+\frac{5}{3}\zeta_2
+\frac{17}{96}\right) + n_f C_F \left(\frac{\zeta_2}{6}
+\frac{7}{48}\right)+n_f \frac{C_A}{6}\right\}
\nonumber \\ && \hspace{-10mm}
{}+\ln\left(\frac{\mu_R^2}{p_T^2}\right) \left\{-\frac{\beta_0}{2}
\left[\frac{3}{4}C_F+C_F\ln\left(\frac{-t}{s}\right)
+C_A\ln\left(\frac{-u}{s}\right)\right]  
\ln\left(\frac{p_T^2}{s}\right)+\frac{\beta_0}{2} T_1^{qg}
+\frac{\beta_1}{16}\right\} \, ,
\eeqa
where
\beq
{\gamma'}_{g/g}^{(2)}=C_A^2\left(\frac{2}{3}+\frac{3}{4}\zeta_3\right)
-n_f\left(\frac{C_F}{8}+\frac{C_A}{6}\right) \, . 
\eeq

For both processes the coefficients of the ${\cal D}_3(s_4)$, 
${\cal D}_2(s_4)$, and ${\cal D}_1(s_4)$ terms were given previously 
in Ref. \cite{KO1}. 
The additional subleading ${\cal D}_0(s_4)$ and
$\delta(s_4)$ terms presented here are new. 

\begin{figure}
\includegraphics[height=4 in]{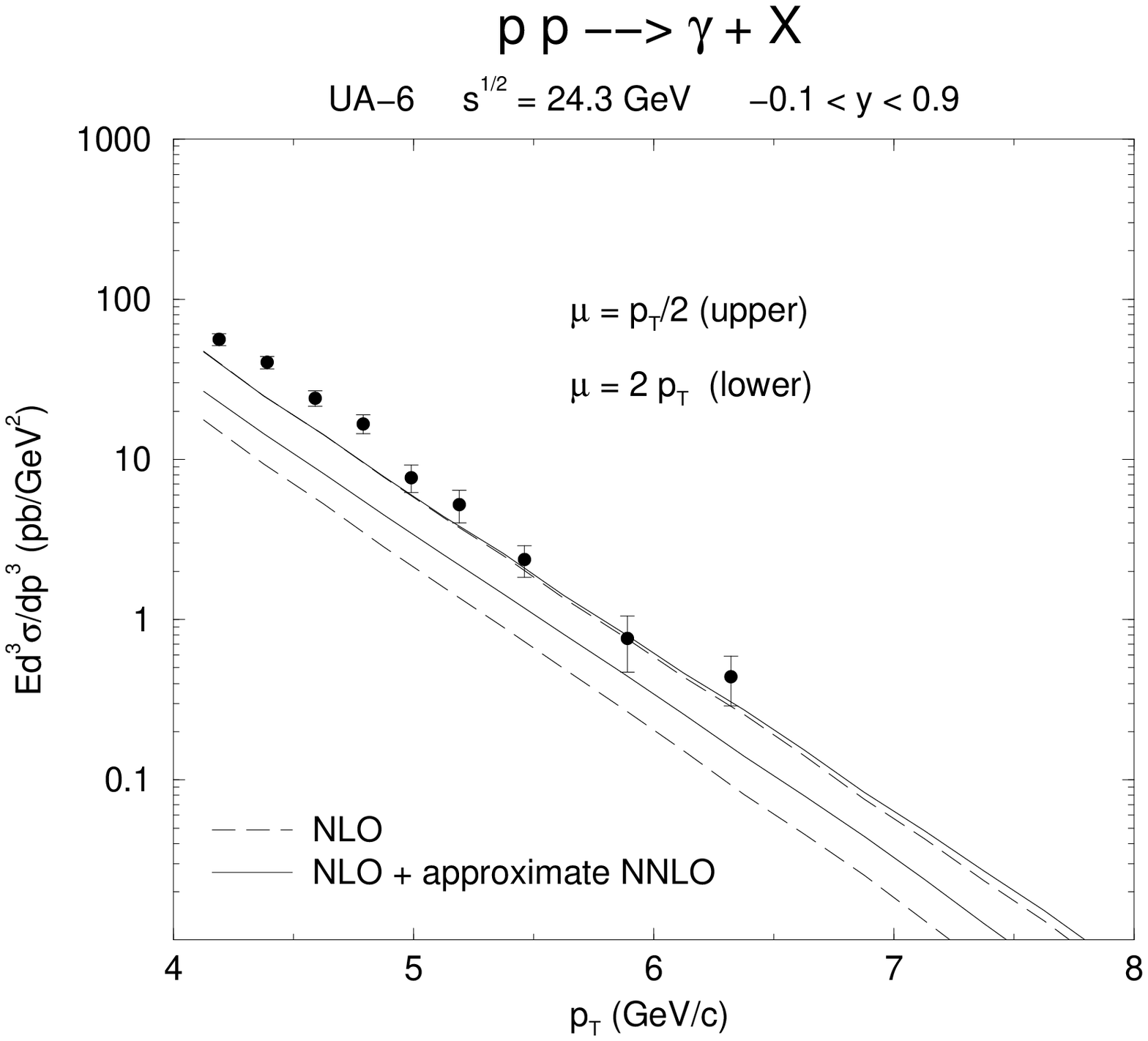}
\caption{NLO and NNLO results for direct photon production
in hadronic collisions compared to $pp$ data from the UA-6 Collaboration 
\cite{UA6} at $\sqrt s = 24.3\ $ GeV.}
\label{fig1}
\end{figure}

\begin{figure}
\includegraphics[height=4 in]{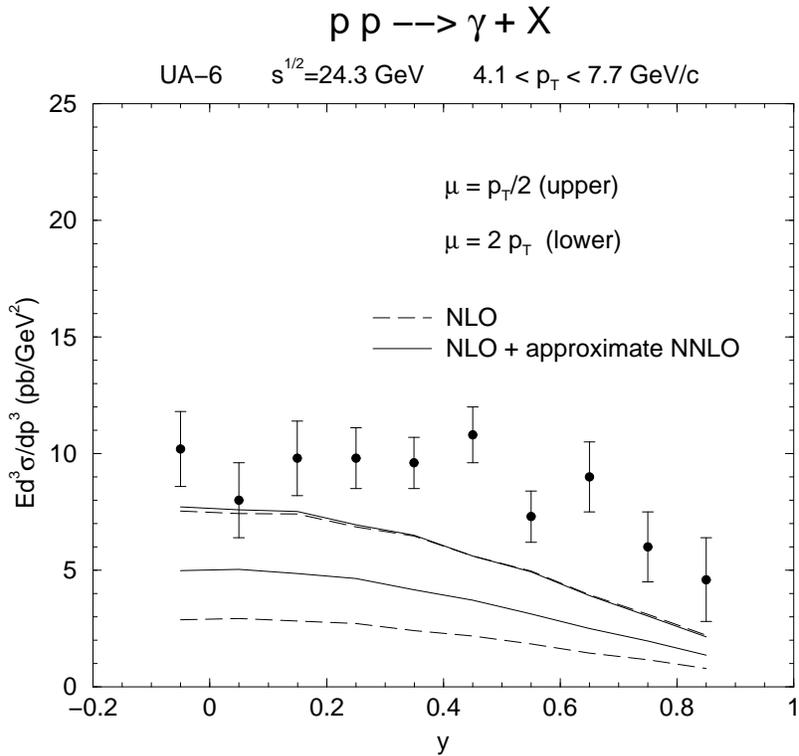}
\caption{NLO and NNLO results for direct photon production
in hadronic collisions compared to $pp$ data for the rapidity distribution 
from the UA-6 Collaboration 
\cite{UA6} at $\sqrt s = 24.3\ $ GeV.}
\label{fig2}
\end{figure}

\begin{figure}
\includegraphics[height=4 in]{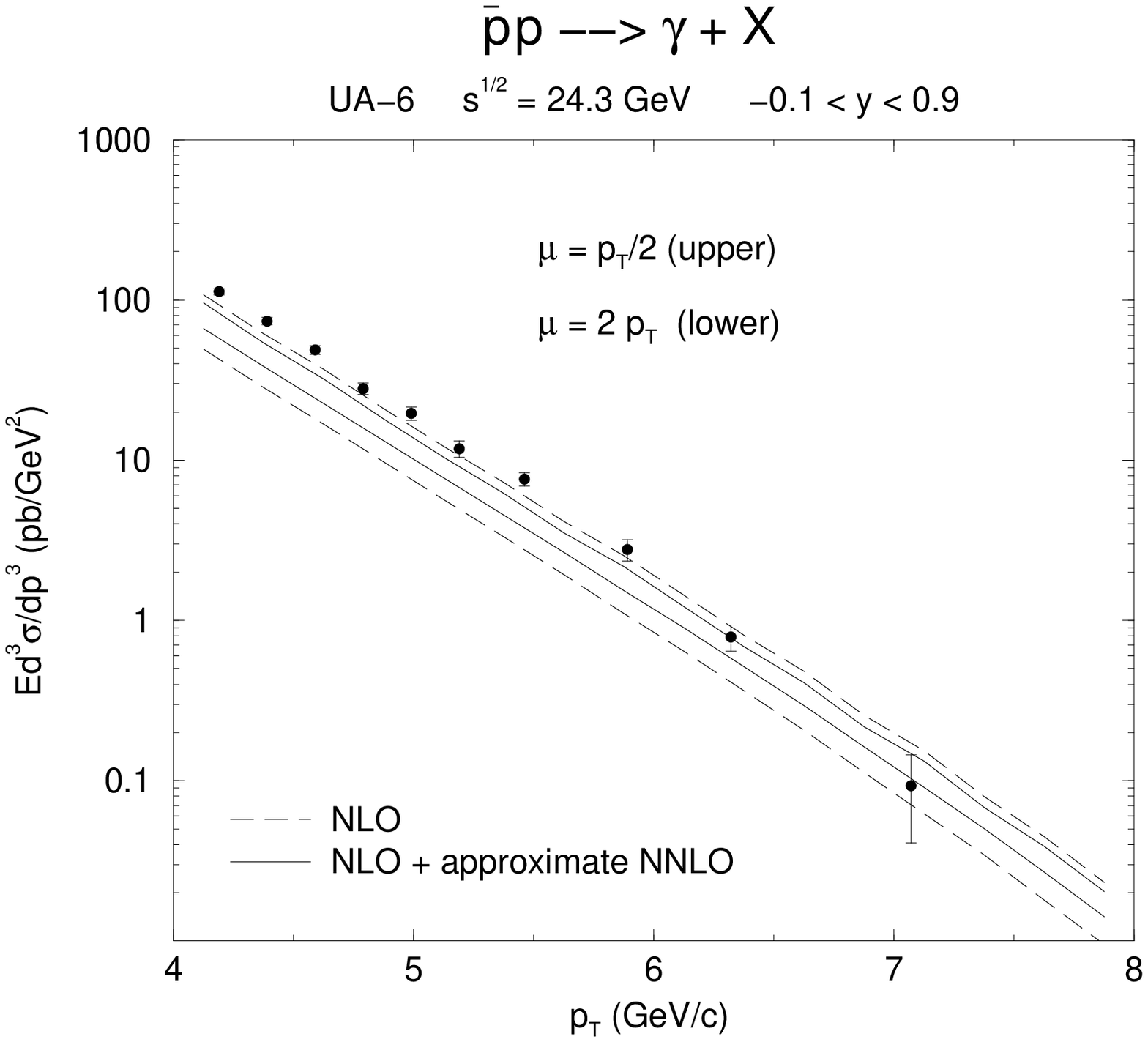}
\caption{NLO and NNLO results for direct photon production
in hadronic collisions compared to $p \overline p$ data from the 
UA-6 Collaboration 
\cite{UA6} at $\sqrt s = 24.3\ $ GeV.}
\label{fig3}
\end{figure}

\begin{figure}
\includegraphics[height=4 in]{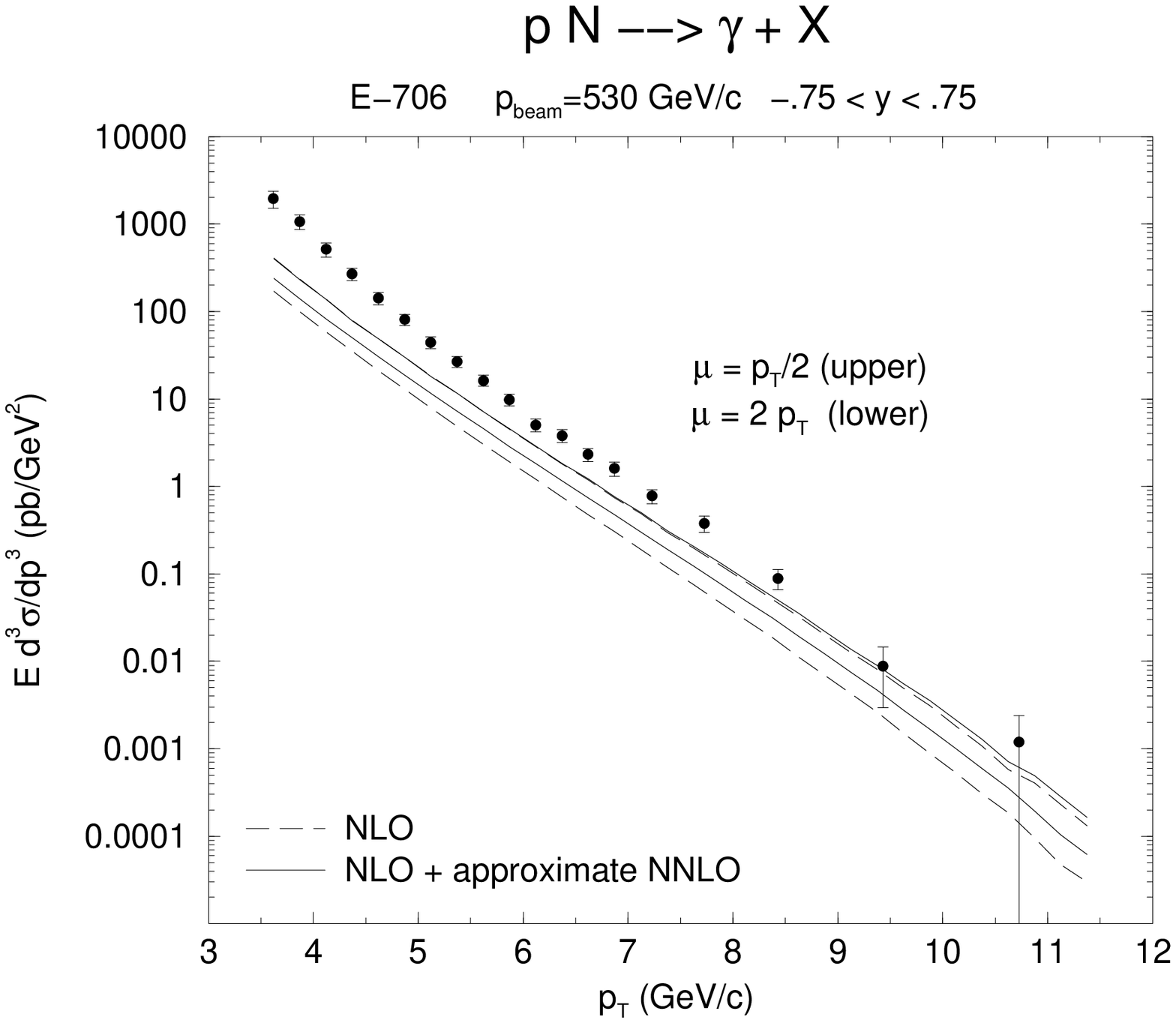}
\caption{NLO and NNLO results for direct photon production
in hadronic collisions compared to data from the E-706 Collaboration 
\cite{E706} at p$_{lab} = 530\ $ GeV/c.}
\label{fig4}
\end{figure}

\section{Numerical results}
\label{sec:results}

In order to show the effect of including the new NNLO terms, the same 
procedure employed in Ref. \cite{KO1} has been used. First, a complete 
NLO calculation of the appropriate cross section is performed using a 
program \cite{BOO} which employs the phase-space slicing technique 
described in Ref. \cite{HO}. The original NLO calculation has been extended 
to include a complete NLO treatment of the bremsstrahlung contribution. 
The Set 2 fragmentation functions of \cite{BFGW} have been used along with the 
CTEQ6M parton distribution functions \cite{CTEQ6}. In all cases the 
factorization and renormalization scales have been set equal to a common 
scale $\mu$ which has been chosen to be proportional to the photon's 
transverse momentum.
Once the NLO results have been obtained, the approximate NNLO contributions 
can be added to them. Several examples are discussed below, comparing the 
NLO and NLO + approximate NNLO results.
 
In Fig. \ref{fig1} the NLO (dashed curves) and NLO plus approximate NNLO 
results (solid curves) are compared to data 
\cite{UA6} from the UA-6 Collaboration for proton proton interactions. In 
each case the upper (lower) curve corresponds to the scale choice of $\mu 
= p_T/2$ $(2 p_T)$. 
The pattern demonstrated previously in Fig. 6 of Ref. \cite{KO1} is still 
found to be true, even after the addition of the newly calculated terms 
presented in this work. The scale dependence of the NLO result is greatly 
decreased 
by the addition of the NNLO terms. Furthermore, one can see that for the 
scale choice of $\mu=p_T/2$ the NNLO scale dependent terms give a negligible 
contribution. Even with the NNLO contributions, the results lie somewhat 
below the data at the lower values of $p_T$.

In Fig. \ref{fig2} the rapidity dependence is shown for the UA-6 proton 
proton data. Again, the scale dependent NNLO terms give a negligible 
contribution for the choice $\mu=p_T/2$ and the overall scale dependence 
is greatly reduced when the NNLO terms are added. As noted previously, the 
curves lie below the data over the majority of the rapidity range shown. 
Of course, this distribution is dominated by the contributions from the 
low end of the $p_T$ range, so this is no surprise, given the results shown in 
Fig. \ref{fig1}. 

In Fig. \ref{fig3} the photon $p_T$ distribution is shown for the case of 
$p \overline p$ interactions and compared to data from the UA-6 Collaboration 
\cite{UA6}. Whereas the $pp$ reaction is dominated by the 
$q g \rightarrow \gamma q$ subprocess, the $p \overline p$ reaction receives 
additional significant contributions from the $q \overline q \rightarrow 
\gamma g$ subprocess. Nevertheless, a pattern similar to that in the previous 
two figures is apparent here as well. Note, however, that for the 
case of $\mu = p_T/2$, the NNLO contribution is negative, further 
reducing the scale dependence shown in Fig. \ref{fig3}. As for the $pp$ case, 
the band formed by  the theoretical curves lies somewhat below the data
at the lower values of $p_T$. 

Finally, in Fig. \ref{fig4} a similar comparison is made to data from the 
E-706 Collaboration \cite{E706}. The same behavior seen in the previous 
figures is evident here, although the theoretical band now lies significantly 
below the data at the lower end of the range covered by the data. 

\section{Conclusions}
\label{sec:conclusions}

We have extended the results of Ref. \cite{KO1} to include additional NNLO 
subleading logarithms resulting from soft gluon corrections for direct photon 
production. The additional terms are numerically small and the results remain 
qualitatively the same as in the previous analysis. In particular, the 
reduced scale dependence relative to NLO calculations remains.

\begin{acknowledgments}
The research of N.K. has been supported by a Marie Curie Fellowship of 
the European Community programme ``Improving Human Research Potential'' 
under contract number HPMF-CT-2001-01221.
The research of J.O. is supported in part by the U.S. Department of Energy.
\end{acknowledgments}

\end{document}